\documentclass[lettersize,journal]{IEEEtran}
\usepackage{amsmath,amsfonts}
\usepackage{algorithmic}
\usepackage{algorithm}
\usepackage{array}
\usepackage[caption=false,font=normalsize,labelfont=sf,textfont=sf]{subfig}
\usepackage{textcomp}
\usepackage{stfloats}
\usepackage{url}
\usepackage{adjustbox}
\usepackage{verbatim}
\usepackage{threeparttable}
\usepackage{graphicx}
\usepackage{cite}
\usepackage{pifont}
\usepackage{makecell}
\usepackage{booktabs}
\usepackage[table]{xcolor}
\usepackage{colortbl}
\usepackage{color}
\usepackage{multicol}
\usepackage{multirow}
\usepackage{soul}
\begin{document}

\title{Towards Edge General Intelligence via Large Language Models: Opportunities and Challenges}

\author{Handi~Chen$^{\dagger}$, Weipeng~Deng$^{\dagger}$, Shuo~Yang, Jinfeng~Xu, Zhihan~Jiang~\IEEEmembership{Student Member,~IEEE}, \\
Edith~C.H.~Ngai$^{*}$,~\IEEEmembership{Senior Member,~IEEE}, Jiangchuan~Liu,~\IEEEmembership{Fellow,~IEEE}, and Xue~Liu,~\IEEEmembership{Fellow,~IEEE}

\thanks{H. Chen, W. Deng, S. Yang, J. Xu, Z. Jiang and E. C.H. Ngai are with the Department of Electrical and Electronic Engineering, the University of Hong Kong, Hong Kong, China (hdchen, dengf330, shuo.yang, jinfeng, zhjiang@connect.hku.hk, chngai@eee.hku.hk).}
\thanks{J. Liu is with the School of Computing Science, Simon Fraser University, British Columbia, Canada (jcliu@sfu.ca).}
\thanks{X. Liu is with the School of Computer Science, McGill University, Montreal, Quebec, Canada (xueliu@cs.mcgill.ca).}
\thanks{$\dagger$ These authors contributed equally to this work.}
\thanks{$*$ The corresponding author.}
}

\markboth{Journal of \LaTeX\ Class Files,~Vol.~14, No.~8, August~2021}%
{Shell \MakeLowercase{\textit{et al.}}: A Sample Article Using IEEEtran.cls for IEEE Journals}

\newcommand{\hd}[1]{\textcolor{yellow}{{#1}}}
\newcommand{\wpd}[1]{\textcolor{red}{{#1}}}
\newcommand{\ys}[1]{\textcolor{olive}{{#1}}}
\newcommand{\xjf}[1]{\textcolor{cyan}{{#1}}}
\maketitle

\begin{abstract}
Edge Intelligence (EI) has been instrumental in delivering real-time, localized services by leveraging the computational capabilities of edge networks. The integration of Large Language Models (LLMs) empowers EI to evolve into the next stage: Edge General Intelligence (EGI), enabling more adaptive and versatile applications that require advanced understanding and reasoning capabilities. However, systematic exploration in this area remains insufficient. This survey delineates the distinctions between EGI and traditional EI, categorizing LLM-empowered EGI into three conceptual systems: centralized, hybrid, and decentralized. For each system, we detail the framework designs and review existing implementations. Furthermore, we evaluate the performance and throughput of various Small Language Models (SLMs) that are more suitable for deployment on edge devices. This survey provides researchers with a comprehensive vision of EGI, offering insights into its vast potential and establishing a foundation for future advancements in this rapidly evolving field.
\end{abstract}


\section{Introduction}

Edge computing has emerged as a crucial network paradigm, processing data closer to its source to reduce latency and resource demands compared to traditional cloud-centric systems. The integration of Artificial Intelligence (AI) into edge devices enables local data analysis, eliminating the need for continuous cloud communication and facilitating devices to make rapid, independent decisions. AI-enhanced Edge Intelligence (EI) advances the adaptability of distributed systems, enabling faster responses in dynamic environments. This evolution unlocks the potential for latency-sensitive applications, transforming industries such as autonomous vehicles, smart homes, industrial automation, and healthcare by introducing more intelligent and responsive technologies.

Tracing the evolution of AI, which aims to emulate human cognitive abilities, we can categorize its development into three stages: narrow, broad, and general intelligence. Following this development trajectory, EI can likewise be divided into Edge Narrow Intelligence (ENI), Edge Broad Intelligence (EBI), and Edge General Intelligence (EGI). 
ENI, empowered by extensive training, excels at performing specific, well-defined tasks, such as facial recognition. EBI transcends individual tasks, enabling systems to manage multiple interconnected functions, such as optimizing traffic flows in smart cities. {Inspired by Artificial General Intelligence (AGI) \mbox{\cite{goertzel2014artificial}}, future edge devices are expected to possess capabilities such as comprehension, reasoning, and generation. EGI enables edge devices to achieve, and in some cases surpass, human-level cognitive abilities across a wide range of tasks.}

\begin{figure*}[!t]
    \centering
    \includegraphics[width=\linewidth]{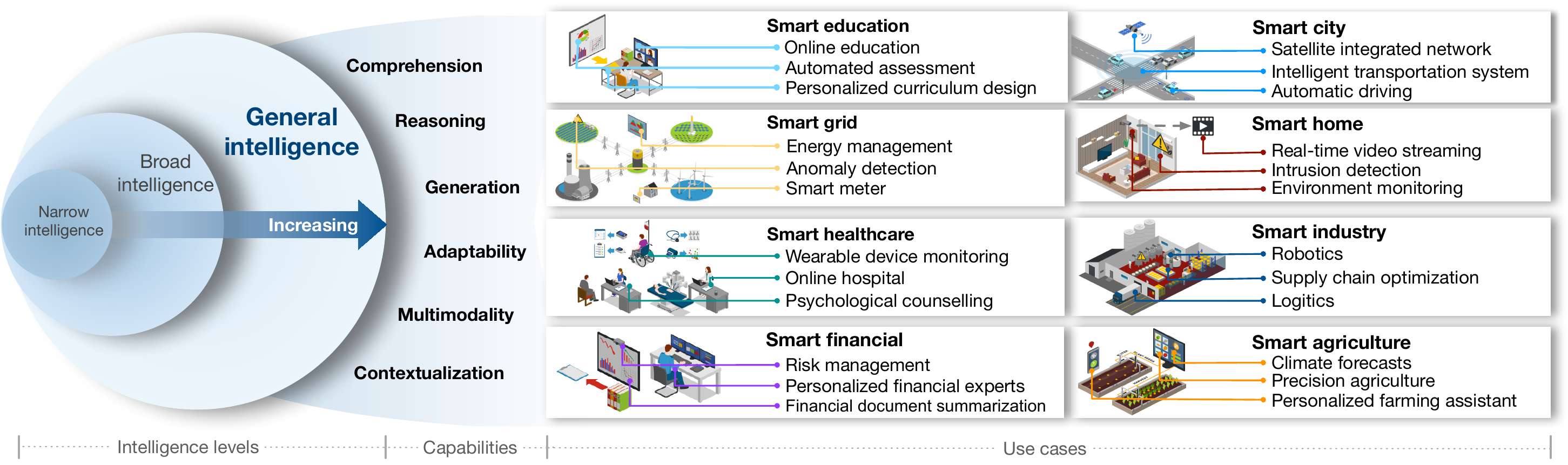}
    \caption{An illustration of increasing intelligence from narrow to general intelligence. The main capabilities of general intelligence include comprehension (e.g., information summarization), reasoning (e.g., logic inference), generation (e.g., writing a story), adaptability (e.g., compatibility with various scenarios), multimodality (e.g., images and audios), and contextualization (e.g. context-aware dialogue), supporting various use cases.}
    \label{fig:archi}
\end{figure*}

However, achieving EGI presents significant challenges for resource-constrained edge devices, requiring them to autonomously reason, learn, and adapt to dynamic environments. 
{Large Language Models (LLMs)\footnote{LLM: https://openai.com/index/better-language-models} offer a promising solution. LLMs are neural networks with a large number of parameters, trained on an extensive corpus, to understand and generate human language for various Natural Language Processing (NLP) tasks.}
Recent advancements in LLMs,
such as GPT-4\footnote{GPT-4: \url{https://openai.com/index/gpt-4}} and LLaMA\footnote{LLaMA: \url{https://github.com/meta-llama}}. 
These developments lay the foundation for general intelligence \cite{shen2024large}. Leveraging LLMs to achieve EGI opens new avenues by enabling edge devices to handle complex tasks. As shown in Fig. \ref{fig:archi}, this integration paves the way for real-time, context-aware applications in various use cases, such as personalized healthcare, smart assistants, and customer service with edge networks \cite{cui2023llmind}.

To inspire innovation and explore the potential of LLM-empowered EGI, this paper proposes three conceptual EGI systems, offering a comprehensive framework for researchers to integrate LLMs into the edge computing ecosystem. We begin by outlining the three levels of EI, followed by a systematic review of recent advancements in LLM-enhanced EGI. Upon this basis, we introduce three system architecture designs, including centralized, hybrid, and decentralized EGI, accompanied by a review of their implementation strategies and an analysis of their respective advantages, disadvantages, and feasibility. We finally conclude by discussing future directions, highlighting the transformative opportunities this survey holds for the evolution of EGI.

\section{Background of EI and LLMs}
In this section, we provide a brief overview of the background knowledge about EI and LLMs.
\subsection{Edge Computing and EI}
Edge computing proposes a paradigm shift by relocating computational resources from distant, centralized cloud servers to edge devices closer to end-users. These devices may include smartphones, industrial machines, sensors, video cameras, or any other data collection devices. Edge computing is designed to enhance the Quality of Service (QoS) for heterogeneous mobile devices and infrastructures in wireless networks by mitigating substantial transmission latency, bandwidth limitations, and connectivity issues caused by cloud computing. It facilitates faster and more efficient data processing while enhancing privacy and security measures. For example, smart traffic light control powered by edge computing processes sensor data in real-time to manage traffic flow more efficiently. Instead of sending data to a central server for processing, the computation is done locally at the traffic intersection, reducing latency and improving the responsiveness of the traffic management system. 
With advances in AI efficiency, the increasing number of IoT devices, and the ascendancy of edge computing, the potential of EI has now been actualized. The evolution of AI is driving the future of edge computing towards EI. 


\subsection{Large Language Models}
Recent advancements in LLMs have demonstrated a spectrum of impressive emergent abilities, leading to significant paradigm shifts in AI. These models display an exceptional capacity for understanding human instructions and demonstrating cognitive capabilities that closely mirror human thinking, paving the way for numerous opportunities across various fields. LLMs have made significant strides in complex task planning and reasoning, skills indispensable for problem-solving and decision-making.
For instance, ChatGPT, developed by OpenAI, excels in general problem-solving, assisting users with a variety of tasks, including solving mathematical problems, devising travel plans, and analyzing investment data to guide decisions. 
As LLMs continue to advance toward general AI, they demonstrate an increasing ability to generate complex, contextually relevant responses across diverse domains, aligning more closely with human-like reasoning.

There has been a surge in developing domain-specific {intelligent systems} based on generic LLMs to integrate expert knowledge across various fields. Retrieval-Augmented Generation (RAG) has emerged as a powerful method, enabling LLMs to dynamically access external knowledge bases during response generation. By integrating relevant information from diverse sources, RAG significantly enhances the accuracy and contextual relevance of outputs, particularly for specialized tasks. 
{Furthermore, to deploy LLMs in domain-specific systems, fine-tuning techniques are employed to augment generic models with domain-specific knowledge while preserving their general knowledge for handling multiple tasks.}
This strategy has proven effective in models such as MedicalGPT\footnote{MedicalGPT: \url{https://github.com/shibing624/MedicalGPT}} and FinGPT\footnote{FinGPT: \url{https://github.com/AI4Finance-Foundation/FinGPT}}, which serve as valuable tools in the medical and financial sectors {while retaining the general capabilities of LLMs.}

\section{LLMs: Evaluating the Next Level of EI}


In this section, we first provide a concise definition of general intelligence to delineate it from traditional intelligence. Following this, we discuss the motivations driving the integration of general intelligence into edge computing, highlighting the potential benefits and transformative impacts on EGI systems. 

\subsection{Hierarchical Cognitive Abilities of EI}

Human intelligence encompasses the cognitive ability to learn, reason, solve problems, adapt to new situations, and understand complex ideas. With this as the ultimate goal, AI systems can be structured hierarchically, including three levels of cognitive ability: narrow, broad, and general intelligence, as shown in Fig. \ref{fig:archi}. The explanations of these three-level intelligence are outlined below:
 
\begin{itemize}
    \item Narrow Intelligence: intelligent systems designed to perform specific tasks or solve particular problems.

    \item Broad Intelligence: intelligent systems that can perform a wider range of tasks across different domains but still lack the full versatility of human-like understanding.

    \item General Intelligence: intelligent systems that possess the ability to understand, learn, and apply knowledge across a wide range of tasks, similar to human intelligence.
\end{itemize}

 
 

The development of EI is closely linked to advancements in AI
to support diverse applications in areas like smart transportation and industrial automation. 
To be specific, the initial level of EI implements narrow intelligence, focusing on specific tasks. For example, ENI-enabled edge devices can analyze network traffic to detect unusual patterns or potential security threats. These models excel in their specialized tasks due to extensive, task-specific training. 

Expanding from narrow capabilities, edge devices with broad intelligence are able to handle a range of interconnected tasks within a domain, constructing a more versatile and intelligent system. An intelligent transportation system serves as an EBI example, where edge devices integrate various functions—such as monitoring traffic flow, controlling traffic lights, and detecting congestion—to optimize traffic patterns and enhance efficiency.

{However, traditional EI, whether ENI or EBI, remains limited to handle specific tasks or domains with predifined workflows or components, such as ADMarker \mbox{\cite{ouyang2024admarker}}, and 
ACSIS \mbox{\cite{valacheas2023acsis}}.}
In contrast, {EGI is a type of intelligence empowers edge devices to achieve or surpass human-level cognitive abilities across a wide range of tasks.}
{As shown in Fig. 1, EGI systems leverage cross-domain knowledge to adapt to diverse tasks and environments, offering versatile solutions for complex problems. Their architectures are highly flexible and integrated, enabling EGI systems to efficiently manage diverse data and tasks by seamlessly combining components from multiple domains.} 
\begin{figure*}[!t]
\centering
\subfloat[]{\includegraphics[width=0.3\textwidth]{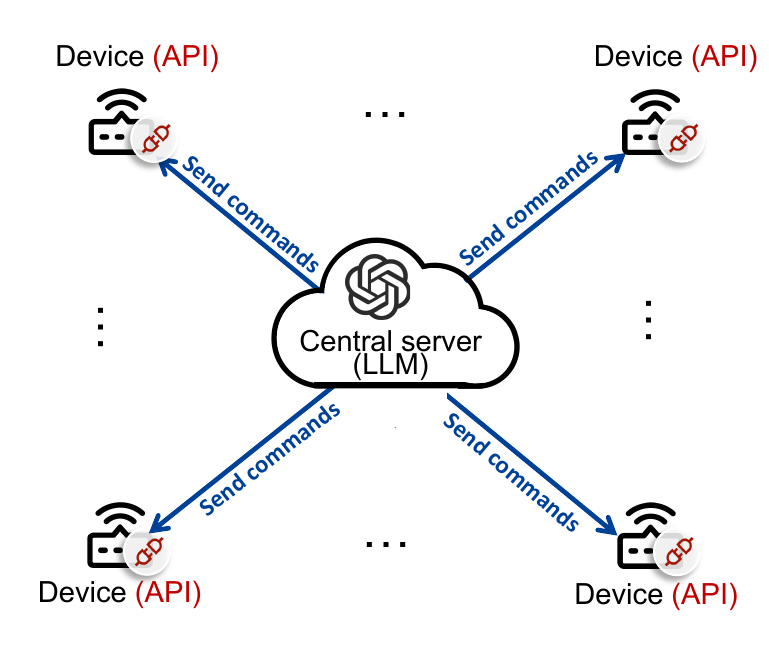}%
\label{fig_archi_1}}
\hfil
\subfloat[]{\includegraphics[width=0.3\textwidth]{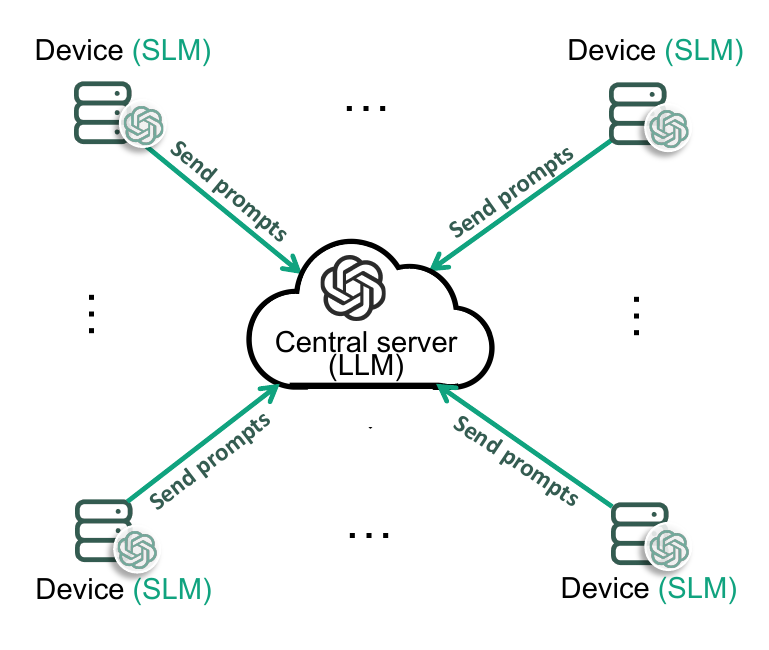}%
\label{fig_archi_2}}
\hfil
\subfloat[]{\includegraphics[width=0.3\textwidth]{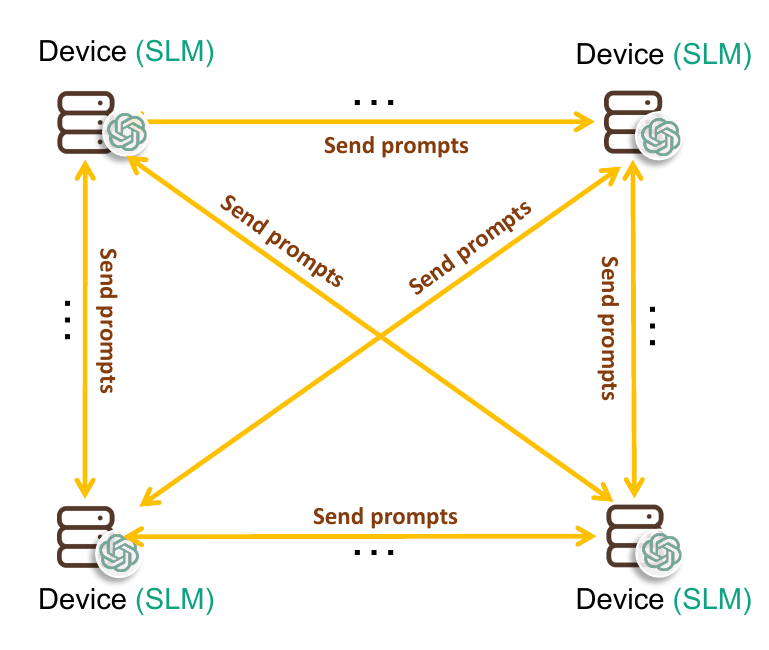}%
\label{fig_archi_3}}
\caption{Architectures of centralized, hybrid, and decentralized EGI systems (LLM and SLM represent large language model and small language model, respectively). (a) Centralized EGI system. (b) Hybrid EGI system. (c) Decentralized EGI system.}
\label{fig_sim}
\end{figure*}

\subsection{LLM-enhanced EGI}

LLMs have revolutionized various fields by performing tasks that previously required human intelligence, such as coding assistance (e.g., Codex and Copilot), math problem-solving, travel planning, robot operations, and complex reasoning. Their ability to learn from diverse data sources grants them a level of general intelligence, making them valuable in dynamic real-world settings. For instance, LLMs can recognize that both a factory worker and a motorcyclist without helmets violate safety regulations, thus eliminating the need for additional data collection. In summary, Fig. \ref{fig:archi} illustrates the six primary capabilities of LLMs essential for constructing EGI-enhanced edge computing.

{Compared to traditional EI systems that use smaller neural networks like CNNs, and ResNet, LLM-enhanced EGI systems typically incur higher computational costs. 
Although optimization techniques like quantization and pruning can reduce computational demands, resource overload remains a significant challenge for EGI system deployment.}
Moreover, LLMs' adaptability and compatibility with various edge sensors and hardware help address deployment challenges, enabling effective coordination of edge devices in complex environments. Enhanced by LLMs, EGI offers more flexible user interactions to manage complex tasks, leveraging strengths in NLP, task generalization, and adaptability. Fig. \ref{fig:archi} presents a vision of several promising real-world use cases for LLM-enhanced EGI. 
Furthermore, LLMs mitigate the limitations of traditional EI, which often requires specific data for distinct tasks and scenarios, thereby reducing computational cost for data processing and enhancing privacy and security. 

\section{Edge General Intelligence Systems}

\begin{table*}[t]
\caption{Overview of representative LLM-based EGI systems across centralized, hybrid, and decentralized architectures.}
\centering
\renewcommand\arraystretch{1.2}
\scalebox{0.9}{\begin{tabular}{p{0.5cm}p{0.4cm}p{1cm}p{1.8cm}p{5.8cm}p{1.9cm}p{1.5cm}p{1.9cm}p{1.5cm}}
\toprule
\multirow{2}{*}{\textbf{Archi.}} & \multirow{2}{*}{\textbf{Ref.}} & \multirow{2}{*}{\textbf{Senarios}} &  \multirow{2}{*}{\textbf{Datasets}} & \multirow{2}{*}{\textbf{Description}}  & \multicolumn{2}{c}{\textbf{Cloud}} & \multicolumn{2}{c}{\textbf{Edge}}\\
\cline{6-7}\cline{8-9}
   &  &  & & & \textbf{Model} & \textbf{Tasks} & \textbf{Model} & \textbf{Tasks} \\
\midrule

\multirow{5}{*}[-43pt]{\rotatebox{90}{\textit{Centralized}}} &
\cite{yang2024llm} & Digital twin & - & Use LLM to simulate a population group's arrivals and distribution across stores for optimization. & GPT-3.5, GPT-4 & Inference & - & - \\

& \cellcolor{gray!20}\cite{kannan2023smart} & \cellcolor{gray!20}Robotic & \cellcolor{gray!20}Benchmark dataset& \cellcolor{gray!20}Use LLM to convert high-level task instructions into a multi-robot task plan. & \cellcolor{gray!20}GPT-4, GPT-3.5, Llama-2 (70B), Claude-3-Opus & \cellcolor{gray!20}Inference & \cellcolor{gray!20}- & \cellcolor{gray!20}- \\

& \cite{shen2024large} & Edge network & User request dataset& Use GPT to interpret user intentions and generate code in a cloud-edge-client framework. & GPT-3 (350M, 6.7B, 175B), GPT-3 IT (175B) & Inference, fine-tuning & - & - \\


& \cellcolor{gray!20}\cite{cui2023llmind} & \cellcolor{gray!20}IoT & \cellcolor{gray!20}- & \cellcolor{gray!20}Use LLM to transform high-level verbal instructions into a control script. & \cellcolor{gray!20}ChatGPT & \cellcolor{gray!20}Inference & \cellcolor{gray!20}- & \cellcolor{gray!20}- \\

& \cite{yuan2024mobile} & Mobile device & - & Propose M4, manage a foundation model serving diverse tasks via lightweight adapters in mobile AI. & RoBERTa, BERT, DistilBERT and so on & Inference, fine-tuning & - & - \\
\midrule

\multirow{3}{*}[-35pt]{\rotatebox{90}{\textit{Hybrid}}} & \cellcolor{gray!20}\cite{ding2024hybrid} & \cellcolor{gray!20}IoT  &  \cellcolor{gray!20}MixInstruct & 
\cellcolor{gray!20}Use a {query scheduler, named ``router"}, to assign queries to SLMs or LLMs based on predicted difficulty and desired quality. & \cellcolor{gray!20}GPT-3.5-turbo, Llama-2 (13B) & \cellcolor{gray!20}Inference & \cellcolor{gray!20}FLAN-T5 (800M), Llama-2 (7B,13B) & \cellcolor{gray!20}Inference\\



& \cite{zhang2024edgeshard} & Edge network  & WikiText-2 & Propose a collaborative edge computing framework using dynamic programming to optimize inference latency and throughput. & Llama-2 (7B, 13B,70B) & Inference & Llama-2 (7B, 13B,70B) & Inference \\

& \cellcolor{gray!20}\cite{hao2024hybrid} & \cellcolor{gray!20}Edge network & \cellcolor{gray!20}GSM8K, HumanEval, NaturalQuestion & \cellcolor{gray!20}Combine SLMs on edge devices with cloud-based LLMs to reduce costs and optimize task performance. & \cellcolor{gray!20}Llama-2 (7B) & \cellcolor{gray!20}Inference & \cellcolor{gray!20}TinyLlama (1.1B) & \cellcolor{gray!20}Inference \\

\midrule

\multirow{4}{*}[-35pt]{\rotatebox{90}{\textit{Decentralized}}} & \cite{pan2024agentcoord} & Edge network & - & Present a visual framework to explore coordination strategies in multi-agent collaboration. & - & - & GPT-4, Mistral (8*7B) & Inference, fine-tuning\\



& \cellcolor{gray!20}\cite{jiang2023large} & \cellcolor{gray!20}6G networks & \cellcolor{gray!20}Cornell Movie-Dialogs Corpus dataset & \cellcolor{gray!20}Propose an LLM-enhanced multi-agent system with tailored communication tools to optimize task-solving through collaboration in 6G networks. & \cellcolor{gray!20}- & \cellcolor{gray!20}- & \cellcolor{gray!20}GPT-3.5 & \cellcolor{gray!20}Inference \\

& \cite{yu2024edge} & Edge network & WikiText, MMLU & Propose an Edge-LLM framework, featuring layer-wise compression, adaptive tuning, and hardware scheduling for optimal performance. & - & - & Llama (7B) & Inference \\

& \cellcolor{gray!20}\cite{zhang2024building} & \cellcolor{gray!20}IoT & \cellcolor{gray!20}TDW-House & \cellcolor{gray!20}Introduce CoELA, a multi-agent cooperation framework that leverages LLMs for decentralized control in complex environments. & \cellcolor{gray!20}- & \cellcolor{gray!20}- &  \cellcolor{gray!20}GPT-4, Llama-2 (13b) & \cellcolor{gray!20}Inference, fine-tuning \\
\bottomrule
\end{tabular}}
\label{tab:ref}
\end{table*}


In this section, we elaborate on the integration of LLMs into edge computing to achieve EGI through three conceptual architectures: centralized, hybrid, and decentralized systems. We provide a comprehensive analysis of these architectures, detailing their frameworks, implementation strategies, and key discussions. Table \ref{tab:ref} presents several representative EGI systems.

\subsection{Centralized EGI System}
\subsubsection{System Framework}

Due to the extensive training requirements and large parameter scales of LLMs, significant computational resources are necessary for both training and inference tasks. To effectively implement an EGI system, a straightforward approach is to deploy LLMs in a centralized cloud server, illustrated in Fig. \ref{fig:archi}(a). This configuration allows the system to fully leverage the powerful computational resources available in the central server.

{In the centralized framework, LLMs are typically hosted on powerful central servers with abundant computational resources, while edge devices serve as tools accessible by the centralized LLM.} By analyzing data collected from each edge device, the central server gains insights into various scenarios and tasks. Each edge device provides an Application Programming Interface (API), enabling the LLM to generate code that integrates and utilizes these devices collectively to complete tasks. {Furthermore, leveraging Key-Value cache (KV-cache) enhances the response efficiency by storing intermediate attention outputs from previous tokens, enabling the central server to reuse these results and improving concurrency during sequential processing.} This approach fosters a scalable and adaptable system, where adding more edge devices extends the LLM's functionality, thereby improving its efficiency in managing and executing tasks.

\subsubsection{Implementation}
Several studies demonstrate practical implementations of this framework. Yang \textit{et al.}~\cite{yang2024llm} propose an LLM-driven digital twin using GPT to simulate customer behaviors and preferences in multi-store malls, enabling reinforcement learning-based optimization to adjust environmental conditions, such as temperature. Kannan \textit{et al.}~\cite{kannan2023smart} introduce SMART-LLM, a centralized system that improves multi-robot task planning through a structured process of task decomposition, coalition formation, task allocation, and task execution, with attention to environmental context and robot capabilities. In smart homes, the Sasha framework~\cite{king2024sasha} processes loosely defined commands (e.g., ``make it cozy.") into JSON format action plans, employing a four-stage task planning methodology similar to SMART-LLM. Shen \textit{et al.}~\cite{shen2024large} integrate language understanding, planning, and code generation capabilities with federated learning to coordinate edge models for diverse user requirements. Furthermore, LLMind~\cite{cui2023llmind} utilizes LLMs as central orchestrators, enhancing the coordination of domain-specific AI modules and IoT devices to execute complex tasks with improved performance beyond LLMs' general knowledge.

\subsubsection{Discussion}
\paragraph{Advantages}
Among the three proposed architectures, the centralized EGI system is the most cost-effective to construct and scale. By leveraging the high processing power of a central server, edge devices act as extensions with minimal additional power consumption. This approach allows the server, with its comprehensive network view, to analyze global data from all edge devices and make optimal decisions. Consequently, the system efficiently coordinates tasks and resources, enhancing overall performance.

\paragraph{Disadvantages}
The centralized EGI system also has notable disadvantages. As the network's intelligence core, the central server faces a heavy workload managing all edge devices. Simultaneous data transmissions from multiple devices can cause network congestion, limiting scalability. For example, an hour of 1080p video (2.25–3.6 GB) transmitted by thousands of urban cameras could overwhelm the network, resulting in data loss and degraded performance. Privacy and security are also critical concerns. A breach of the central server could grant attackers access to all data from connected edge devices, creating a single point of failure that could disrupt functionality, compromise critical services, and cripple the entire network.

\paragraph{Feasibility Discussion}
Recent advancements in LLMs have improved the praticality of centralized EGI systems. Major corporations now offer ready-to-use APIs with advanced capabilities, simplifying the integration of LLMs into existing edge systems. Upgrading these edge systems often requires only deploying the appropriate APIs.
 
Studies on LLM-based agents have demonstrated that employing LLMs as central schedulers effectively coordinates collaboration among multiple tools~\cite{kannan2023smart, king2024sasha, shen2024large}, enabling complex task completion previously required human intervention. 
{In parallel, multimodal Vision-Language Models (VLMs) have emerged, enabling LLM-based agents to process both visual and textual inputs, thereby expanding their capacity for tasks like image and document interpretation and facilitating more complex decision-making.}
Moreover, research into function calling and tool utilization shows that LLMs can autonomously select and use external tools as needed. This multimodality not only enhances the efficiency of client interactions but also enables comprehensive collaboration among diverse tools for a single task. Table ~\ref{tab:abilit} presents a list of API providers and their abilities, illustrating the diverse options available for organizations looking to implement centralized EGI systems.

\begin{table}[!t]
\centering
\begin{threeparttable}
\renewcommand\arraystretch{1.2}
\caption{LLM API providers and the supported modalities.}
\label{tab:abilit}
\begin{tabular}{p{2.0cm}p{1.1cm}<{\centering}p{1.1cm}<{\centering}p{1.1cm}<{\centering}p{1.1cm}<{\centering}}
\toprule
\textbf{Provider}  & \textbf{Language} & \textbf{Tool} & \textbf{Vision} & \textbf{Audio} \\
\midrule
OpenAI\tnote{1}    & \ding{51}      & \ding{51}  & \ding{51} & \ding{51}   \\
\rowcolor{gray!20} Anthropic\tnote{2} & \ding{51}      & \ding{55}   & \ding{55}   & \ding{55}  \\
Cohere\tnote{3} & \ding{51} & \ding{55} & \ding{55} & \ding{55}\\
\rowcolor{gray!20}  Google\tnote{4}    & \ding{51}      & \ding{51}  & \ding{51}  & \ding{51}  \\ 
Microsoft Azure\tnote{5} & \ding{51} & \ding{51} & \ding{51} & \ding{51}\\
\rowcolor{gray!20} Mistral AI\tnote{6} & \ding{51} & \ding{55} & \ding{55} & \ding{55}\\
\bottomrule

\end{tabular}
\begin{tablenotes}
\footnotesize
\item[1] OpenAI: \url{https://openai.com/api/}
\item[2] Anthropic: \url{https://docs.anthropic.com}
\item[3] Cohere: \url{https://docs.cohere.com/reference/about}
\item[4] Google: \url{https://ai.google/}
\item[5] Microsoft Azure:
\url{https://learn.microsoft.com/en-us/azure/}
\item[6] Mistral AI: \url{https://docs.mistral.ai/api/}
\end{tablenotes}
\end{threeparttable}
\end{table}

\subsection{Hybrid EGI System}
\subsubsection{System Framework}
\begin{table*}[!ht]
\caption{The description of representative SLMs.}
\renewcommand\arraystretch{1.2}
\label{tab:throughput}
\centering
\begin{tabular}{p{2.5cm}p{1.7cm}p{1.7cm}>{\centering\arraybackslash}p{1.5cm}>{\centering\arraybackslash}p{1.5cm}>{\centering\arraybackslash}p{1.5cm}>{\centering\arraybackslash}p{1.5cm}>{\centering\arraybackslash}p{1.7cm}}
\toprule

\multirow{2}{*}{\textbf{Model}} & \multirow{2}{*}{\textbf{Provider}}       & \multirow{2}{*}{\textbf{Language}} & \multirow{2}{*}{\makecell{\textbf{Parameter} \\ \textbf{Size}}} & \multirow{2}{*}{\makecell{ \textbf{Context} \\ \textbf{Length}}} & \multicolumn{2}{c}{\textbf{Throughput$^\ast$ }} & \multirow{2}{*}{\makecell{\textbf{Release} \\ \textbf{Date}}} \\ 

\cline{6-7}
 & & & & & \textbf{requests/s} & \textbf{tokens/s} &  \\ \midrule
Qwen-1.5-0.5B           & Alibaba                          & Multilingual               & 0.5B                             & 32k                              & 34.36           & 14406.54     & 2024.02                        \\


\rowcolor{gray!20}
Tinyllama               & jzhang38                         & English                    & 1.1B                             & 2K                             & 31.64           & 15226.31     & 2024.04                        \\

stablelm-2-1\_6b-chat   & Stability-AI                     & English                    & 1.6B                             & 8K                             & 23.66           & 9929.77      & 2024.04                        \\

\rowcolor{gray!20}
Qwen-1.5-1.8B           & Alibaba                          & Multilingual               & 1.8B                             & 8K                             & 22.84           & 9575.21      & 2024.02                        \\

Gemma                   & Google                           & Multilingual               & 2B                               & 8K                             & 24.84           & 10945.68     & 2023.09                        \\

\rowcolor{gray!20}
phi-2                   & Microsoft & English                    & 2.7B                             & 2K      & 13.87           & 6305.34      & 2023.12                        \\

stablelm-3b-4e1t        & Stability-AI                     & English                    & 3B                               & 4K                             & 13.98           & 6293.27      & 2023.09                        \\

\rowcolor{gray!20}
phi-3-mini              & Microsoft & English                    & 3.8B                             & 4K\&128K                         & 9.72            & 4675.83      & 2024.04                        \\

Qwen-1.5-4B             & Alibaba                          & Multilingual               & 4B                               & 32K                              & 12.04           & 5048.83      & 2024.02                        \\

\rowcolor{gray!20}
Qwen-1.5-7B             & Alibaba                          & Multilingual               & 7B                               & 8K                               & 4.66            & 1953.71      & 2024.02                        \\

Mistral-7B              & Mistral AI                       & English                    & 7B                               & 8K                               & 10.18           & 4809.72      & 2023.09                        \\

\rowcolor{gray!20}
Llama-2-7b              & Meta                             & Multilingual               & 7B                               & 4K                               & 4.74            & 2280.35      & 2023.08                        \\

gemma-7b                & Google                           & Multilingual               & 7B                               & 8K                               & 2.71            & 1192.78      & 2024.02                        \\

\rowcolor{gray!20}
phi-3-small             & Microsoft & English                    & 7B                               & 8K                               & 10.82           & 4483.98      & 2024.04                        \\

Llama-3-8B              & Meta                             & Multilingual               & 8B                               & 8K                               & 10.87           & 4496.12      & 2024.04                        \\ \bottomrule
\multicolumn{8}{l}{$\ast$ Throughput measured in one 4090 GPU with 24 GB of memory. The max context window is 4096 for all models.}\
\end{tabular}
\end{table*}

As shown in Fig. \ref{fig:archi}(b), both the central server and edge devices are equipped with language models to enhance intelligence. In a hybrid EGI system, the central server hosts a more advanced and powerful LLM, while the edge devices run Small Language Models (SLMs) due to their limited computational capacity {for simple queries and initial data preprocessing.} Recent advancements in SLMs have enabled such deployments. While these device-side SLMs may not rival the capabilities of cloud-based powerhouses like GPT-4 as demonstrated in the results shown in Table \ref{tab:throughput}. {In hybrid systems, edge-cloud collaborative intelligence requires efficient token-streaming communication protocols to compress and transmit edge-generated responses to balance response latency and performance.}
{Integrating SLMs capable of basic general intelligence tasks (such as NLP and answering common-sense questions) into a mixture framework significantly enhances their functionality. Within this framework, each SLM is fine-tuned as an expert for specific tasks and dynamically selected by a gating mechanism, enabling efficient execution in more complex environments that require direct human interaction, such as smart homes, mobile assistants, and vehicular systems.}

\subsubsection{Implementation}
Hybrid EGI systems provide greater flexibility but introduce deployment complexities. Ding \textit{et al.}~\cite{ding2024hybrid} propose a hybrid LLM architecture that combines models of varying sizes, utilizing a quality-aware {query scheduler, namely the ``router"} to direct queries to the most cost-effective model (large for complex tasks and small for simpler ones), achieving significant cost savings without compromising quality. To enhance edge-cloud cooperation, Yang \textit{et al.} \cite{yang2024edgeFM} propose EdgeFM, a system that leverages Foundation Models (FMs) to enhance the generalization capabilities for on-device models on resource-limited IoT devices. EdgeFM dynamically uploads data to the cloud for FM querying and switches models based on data uncertainty and network conditions, yielding higher accuracy and reduced latency.
Zhang \textit{et al.} \cite{zhang2024edgeshard} present a collaborative framework for LLM inference that partitions models between edge devices and cloud servers, incorporating adaptive device selection and dynamic programming to minimize latency and maximize throughput. Hao \textit{et al.} \cite{hao2024hybrid} propose a dynamic token-level edge-cloud collaboration framework, employing SLMs like TinyLlama on edge devices to interact with cloud-based LLMs for high-quality, cost-effective performance. Collectively, these innovations demonstrate the potential of hybrid EGI systems to balance computational demands while enhancing performance and efficiency across various applications.

\subsubsection{Discussion}
\paragraph{Advantages}

Hybrid EGI systems leverage edge devices equipped with SLMs to enable low-latency decision-making, facilitating quick responses to local events without relying on central server processing. This capability is crucial for real-time applications such as smart homes and autonomous vehicles. Additionally, the system improves load balancing by distributing tasks between edge devices and the central server, preventing overload and enhancing overall efficiency and reliability. Its flexibility allows it to handle diverse tasks with varying constraints, as edge devices can preprocess data and identify essential information for transmission. By using SLMs for semantic communication rather than transmitting raw data, the system conserves network resources, reduces communication overhead, and optimizes performance.

\paragraph{Disadvantages}
However, hybrid EGI systems come with notable disadvantages. Deployment costs are higher due to the requirement to set up both a central LLM and edge SLMs. Maintenance costs are also significant, stemming from the complexity of updating, securing, and managing both the edge devices and the central server. Furthermore, since SLMs lack the processing and decision-making capabilities of LLMs, efficient task scheduling and resource allocation become essential to balance QoE with efficiency.

\paragraph{Feasibility Discussion}
Recent advancements in SLMs have enabled their deploying on edge devices, demonstrating that models with as few as 1.1 billion parameters can achieve a substantial level of general intelligence. These models can understand human language, engage in fluent communication, and answer questions that do not necessitate complex reasoning or in-depth world knowledge. Progress in model optimization techniques has further facilitated SLM deployment on edge devices. Techniques such as quantization and efficient inference frameworks, which incorporate optimizations like kernel fusion, have been crucial. These innovations make it feasible to deploy SLMs on consumer-grade hardware, such as smartphones equipped with a Snapdragon 888 processor and 8GB of RAM, achieving impressive performance levels. 

\subsection{Decentralized EGI System}
\subsubsection{System Framework}

In the decentralized EGI framework, each edge device has its own SLM for autonomous decision-making, enabling local data processing and task execution. {Techniques such as model distillation, quantization, and pruning are essential for reducing the model size and computational demands, enabling deployment on resource-constrained edge devices. From hardware perspective, Neural Processing Units (NPUs) in mobile SoCs and desktop processors allow faster processing with lower power consumption.}
As depicted in Fig. \ref{fig:archi}(c), this decentralized architecture significantly reduces the need for human intervention in managing device interactions, thereby increasing the efficiency of decision-making processes. Furthermore, while these devices operate independently, their ability to collaborate with others is essential in a decentralized EGI architecture. This collaborative capability allows them to share insights, improve overall system performance, and address complex tasks that require collective intelligence.

\subsubsection{Implementation}
Although research on decentralized EGI systems is still in its early stages, several studies have demonstrated their ability to outperform single-agent systems by leveraging the collective knowledge and processing power of multiple intelligent agents. 
To facilitate multi-agent coordination, Zhang \textit{et al.} \cite{zhang2024building} propose the Cooperative Embodied Language Agent (CoELA), a modular framework inspired by cognitive principles. CoELA utilizes LLMs for reasoning, language comprehension, and text generation in decentralized control scenarios with high communication costs. Experiments show that GPT-4-driven CoELA outperforms traditional planning methods, with fine-tuning further enhancing its natural language-based human-agent interactions. Chan \textit{et al.} \cite{chan2023chateval} introduce ChatEval, a multi-agent framework employing diverse communication strategies and agent personas to collaboratively evaluate text. By integrating collective intelligence and cognitive synergy, ChatEval aligns more closely with human preferences, improving evaluation accuracy through multi-perspective dialogue. Jiang \textit{et al.} \cite{jiang2023large} propose a multi-agent system consisting of three components: multi-agent data retrieval, multi-agent collaborative planning, and multi-agent evaluation and reflection. These components refine communication knowledge, generate feasible solutions, and evaluate and improve task outcomes,  addressing challenges such as resource constraints and LLM-induced delays. Yu \textit{et al.} \cite{yu2024edge} develop a framework optimizing LLM adaptation on edge devices to mitigate high computational and memory demands. Their approach combines layer-wise compression through adaptive pruning and quantization, adaptive layer tuning with a voting mechanism to reduce backpropagation depth, and a hardware scheduling strategy to manage irregular computation patterns.

\subsubsection{Discussion}
\paragraph{Advantages}
In the decentralized intelligence system, intelligent edge devices communicate directly, enabling collaborative task execution and reducing manual costs. This approach is particularly effective in environments with poor communication conditions, such as oceans and disaster areas, where distributed collaborative intelligence adapts to diverse scenarios and mitigates single points of failure inherent in centralized and hybrid systems. Local data processing minimizes privacy risks associated with data transmission and storage, while the decentralized architecture enhances scalability and collective intelligence. Real-time responses, achieved without central server communication, optimize network efficiency by conserving bandwidth and reducing load, making the system suitable for applications that require speed, scalability, robustness, and privacy. Unlike traditional decentralized edge systems with minimal collaboration or static rules, EGI systems with SLMs enable dynamic decision-making and adaptive collaboration, effectively addressing the complexities of real-world environments—a critical advantage for edge computing.

\paragraph{Disadvantages}

Despite its advantages, decentralized EGI systems face several challenges. Frequent interactions between edge devices complicate trust mechanisms, making it difficult to enforce consistent policies across a decentralized network. Limited computational capacities of device-level SLMs hinder their ability to manage complex tasks compared to centralized LLMs. Additionally, the significant coordination overhead requires sophisticated protocols to efficiently manage peer-to-peer interactions and maintain network performance. The computational demands of running even an SLM often necessitate hardware upgrades for edge devices, increasing deployment costs.

\paragraph{Feasibility Discussion}
The NLP community has started exploring the use of multiple LLMs to simulate group intelligence behaviors, enabling collaboration and discussion among multiple LLM-based agents. This research field is still in its early stages and remains a novel topic for further investigation and development.

\section{Future Directions}
In this section, we outline key future directions for implementing LLM-empowered EGI to fully unlock the potential of LLMs in resource-constrained edge environments. 

\subsection{Efficient SLM Deployment on Resource-limited Edge Devices}
Deploying efficient general intelligence on resource-limited edge devices for EGI is challenging due to the substantial Random Access Memory (RAM) and computational demands of LLMs and SLMs. Solutions can be categorized into algorithmic and hardware-based approaches. Techniques such as quantization, which reduces the precision of model data, show promise in lowering memory and computational costs. However, even optimized models require at least 500MB of RAM and high-end CPUs, like the Snapdragon 888, for effective inference. While low-bit-rate inference is feasible, training models similarly remain difficult, limiting on-device refinement. Split learning/inference offers a partial solution but relies on optimal model partitioning and secure data transmission. On the hardware side, boosting the processing power and memory of edge devices is critical, {especially for remote and rural areas where limited bandwidth necessitates local deployment of language models. In such environments, restricted power supply imposes constraints on the model size, requiring a careful balance between the response performance of deployed SLMs and bandwidth limitation to central servers.} AI-specific processors, including GPUs, FPGAs, and ASICs (e.g., Google’s TPU), alongside system-level optimizations like distributed inference across device clusters, offer potential solutions. Addressing these challenges is crucial for enabling the broader adoption and efficient operation of LLM-empowered EGI systems in edge computing scenarios.

\subsection{Optimizing Latency of Providing LLM-enhanced EGI Services}
Response latency, encompassing both inference and communication delays, serves as a pivotal determinant of QoS in LLM-augmented EGI systems. Inference latency is driven by the hardware's processing capabilities and the efficiency of inference algorithms. Improvements require advancements in computational methods like model optimization and quantization, alongside hardware innovations such as AI accelerators and optimized processors to reduce delays and enhance throughput. Communication latency stems from data transmission delays over the internet. While text-based LLMs handle small data packages, multimodal LLMs, which process larger inputs like images and videos, face greater challenges. For instance, a one-hour 4K video can range from 9GB to 27GB, straining 5G networks with uplink speeds of around 50Mbps. Addressing this requires efficient data compression and transmission techniques, with future 6G advancements potentially easing these limitations and enabling more practical multimodal LLM-based EGI services.

\subsection{Adapting LLMs for Domain-Specific Applications in Edge Networks}
Current LLMs are trained on broad, general-domain data, while edge systems often require domain-specific knowledge. To meet this need, LLMs must be adapted through continued pre-training or fine-tuning on specialized datasets, enhancing their ability to address domain-specific queries. This adaptation is essential for edge computing, where rapid and accurate responses are critical. Additionally, managing time-varying network conditions to avoid negative impacts caused by network fluctuations. Solutions like dynamically adjusting model complexity and offloading heavier tasks to the cloud help maintain low-latency responses while ensuring essential functions remain on edge devices.

\subsection{Security and Privacy Concerns for LLM-enhanced EGI Services}
Deploying LLMs on edge networks poses significant security challenges, including privacy protection and data integrity. 
In centralized EGI systems, differential privacy adds noise to sensitive data but may reduce query clarity. Decentralized systems mitigate exposure by preprocessing data locally with SLMs and sharing features, though complex communication topologies require robust authorization mechanisms. Federated learning and split learning frameworks further enhance privacy by avoiding raw data sharing.
Maintaining LLM integrity is equally critical, particularly against threats such as model tampering, poisoning, and adversarial attacks, where adversaries manipulate the model or its training data to induce erroneous outputs, especially for decentralized EGI systems. 
To mitigate these risks, strategies like adversarial training, robust verification, and cryptographic protocols are promising. Blockchain ensures authenticity in distributed systems, while attribute-based access control and zero-trust architectures prevent unauthorized access. These solutions are vital for secure, reliable LLM deployment in edge networks.

\subsection{Collaborative LLM-enhanced Edge Systems}
LLM-empowered edge devices bring a remarkable ability to understand and adapt to dynamic environments, making them valuable across various applications. However, leveraging multiple LLM-based agents to collaborate and complete tasks \cite{yu2024edge} introduces unique challenges, particularly in resource-constrained extreme environments such as mountainous regions, earthquake zones, and oceans. These scenarios face issues like limited connectivity, low bandwidth, and unreliable power supplies, which complicate the seamless interaction among agents. Moreover, the unpredictable behaviors of LLM-based agents add another layer of complexity to collaboration. Coordinating tasks efficiently requires not only robust communication protocols and resource management but also a system capable of handling these unpredictable responses. Overcoming these challenges is crucial for deploying EGI systems to effectively support remote tasks such as disaster recovery and environmental exploration.

\section{Conclusion}
In this survey, we explored the evolution of LLM-enhanced EGI and differentiated it from traditional EI. The LLM-enhanced EGI systems are categorized into three conceptual designs, namely centralized, hybrid, and decentralized, each reflecting distinct architectures and operational strategies. For each system, this survey summarized the framework designs, reviewed representative research, and discussed the advantages, disadvantages, and feasibility. Moreover, we compared various SLMs based on their characteristics and throughput, offering valuable insights for orchestrating EGI systems. Looking ahead, we summarized promising future directions for EGI in terms of resource efficiency, service optimization, domain adaptation, interaction security, and multi-agent collaboration. This survey presents a comprehensive vision of EGI and outlines key future directions that will contribute to its ongoing advancement.

\bibliographystyle{ieeetr}
\bibliography{Refe}

\vfill

\end{document}